

\font\titlefont = cmr10 scaled\magstep 4
\font\sectionfont = cmr10
\font\littlefont = cmr5

\magnification = 1200

\global\baselineskip = 1.2\baselineskip
\global\parskip = 4pt plus 0.3pt
\global\abovedisplayskip = 18pt plus3pt minus9pt
\global\belowdisplayskip = 18pt plus3pt minus9pt
\global\abovedisplayshortskip = 6pt plus3pt
\global\belowdisplayshortskip = 6pt plus3pt


\def\endignore{}
\def\ignore #1\endignore{}

\newcount\dflag
\dflag = 0


\def\monthname{\ifcase\month
\or Jan \or Feb \or Mar \or Apr \or May \or June%
\or July \or Aug \or Sept \or Oct \or Nov \or Dec
\fi}

\def\timestring{{\count0 = \time%
\divide\count0 by 60%
\count2 = \count0
\count4 = \time%
\multiply\count0 by 60%
\advance\count4 by -\count0
\ifnum\count4 < 10 \toks1 = {0}
\else \toks1 = {} \fi%
\ifnum\count2 < 12 \toks0 = {a.m.}
\else \toks0 = {p.m.}
\advance\count2 by -12%
\fi%
\ifnum\count2 = 0 \count2 = 12 \fi
\number\count2 : \the\toks1 \number\count4%
\thinspace \the\toks0}}



\def\endtitle{}
\def\title#1\endtitle{\vskip.5in\titlefont
\global\baselineskip = 2\baselineskip
#1\vskip.4in
\baselineskip = 0.5\baselineskip\rm}

\def\endauthors{}
\def\authors#1\endauthors{#1}

\def\endabstract{}
\def\abstract#1\endabstract{\vskip .3in%
\centerline{\sectionfont\bf Abstract}%
\vskip .1in
\noindent#1}

\newcount\nsection
\newcount\nsubsection

\def\section#1{\global\advance\nsection by 1
\nsubsection=0
\bigskip\noindent\centerline{\sectionfont \bf \number\nsection.\ #1}
\bigskip\rm\nobreak}

\def\subsection#1{\global\advance\nsubsection by 1
\bigskip\noindent\sectionfont \sl \number\nsection.\number\nsubsection)\
#1\bigskip\rm\nobreak}


\def\appendix#1#2{\bigskip\noindent%
\centerline{\sectionfont \bf Appendix #1.\ #2}
\bigskip\rm\nobreak}


\newcount\nref
\global\nref = 1

\def\ref#1#2{\xdef #1{[\number\nref]}
\ifnum\nref = 1\global\xdef\therefs{\noindent[\number\nref] #2\ }
\else
\global\xdef\oldrefs{\therefs}
\global\xdef\therefs{\oldrefs\vskip.1in\noindent[\number\nref] #2\ }%
\fi%
\global\advance\nref by 1
}

\def\listrefs{\vfill\eject\section{References}\therefs}


\newcount\cflag
\newcount\nequation
\global\nequation = 1
\def\eqlabel{(1)}

\def\nexteqno{\ifnum\cflag = 0
\global\advance\nequation by 1
\fi
\global\cflag = 0
\xdef\eqlabel{(\number\nequation)}}

\def\lasteqno{\global\advance\nequation by -1
\xdef\eqlabel{(\number\nequation)}}

\def\label#1{\xdef #1{(\number\nequation)}
\ifnum\dflag = 1
{\escapechar = -1
\xdef\draftname{\littlefont\string#1}}
\fi}

\def\clabel#1#2{\xdef\eqlabel{(\number\nequation #2)}
\global\cflag = 1
\xdef #1{\eqlabel}
\ifnum\dflag = 1
{\escapechar = -1
\xdef\draftname{\string#1}}
\fi}

\def\cclabel#1#2{\xdef\eqlabel{#2)}
\global\cflag = 1
\xdef #1{\eqlabel}
\ifnum\dflag = 1
{\escapechar = -1
\xdef\draftname{\string#1}}
\fi}


\def\eeq{}

\def\eqnn #1\eeq{$$ #1 $$}

\def\eq #1\eeq{\xdef\draftname{\ }
$$ #1
\eqno{\eqlabel \rlap{\ \draftname}} $$
\nexteqno}




\def\eeol{& \eqlabel \rlap{\ \draftname}
\nexteqno
\xdef\draftname{\ }}

\def\eolnn{\cr
\global\cflag = 0
\xdef\draftname{\ }}


\def\eqa #1\eeq{\xdef\draftname{\ }
$$ \eqalignno{ #1 } $$
\global\cflag = 0}



\def\npb#1#2#3{{\it Nucl. Phys.} {\bf B#1} (19#2) #3}

\def\prd#1#2#3{{\it Phys. Rev.} {\bf D#1} (19#2) #3}
\def\pr#1#2#3{{\it Phys. Rev.} {\bf #1} (19#2) #3}

\def\prl#1#2#3{{\it Phys. Rev. Lett.} {\bf #1} (19#2) #3}

\def\zpc#1#2#3{{\it Zeit. Phys.} {\bf C#1} (19#2) #3}


\global\nulldelimiterspace = 0pt



\def\frac#1#2{{{#1} \over {#2}}\,}  



\def\Dsl{\hbox{/\kern-.6000em\it D}} 
\def\dsl{\hbox{/\kern-.5600em$\partial$}}
\def\pxpsl{\hbox{/\kern-.5600em$p$}}
\def\ssl{\hbox{/\kern-.5600em$s$}}
\def\epssl{\hbox{/\kern-.5600em$\epsilon$}}
\def\delsl{\hbox{/\kern-.7000em$\nabla$}}
\def\lxpsl{\hbox{/\kern-.5600em$l$}}
\def\kxpsl{\hbox{/\kern-.5600em$k$}}
\def\qxpsl{\hbox{/\kern-.5600em$q$}}
\def\sla#1{\raise.15ex\hbox{$/$}\kern-.57em #1}



\def\roughly#1{\mathrel{\raise.3ex\hbox{$#1$\kern-.75em\lower1ex\hbox{$\sim$}}}}












\ref\LWWV{K.J.F. Gaemers and G.J. Gounaris, \zpc{1}{79}{259};
K. Hagiwara, R.D. Peccei, D. Zeppenfeld and K. Hikasa, \npb{282}{87}{253}.}

\ref\gbloop{For references to the literature, see Refs.~5,7.}

\ref\stu{M.E. Peskin and T. Takeuchi, \prl{65}{90}{964};
W.J. Marciano and J.L. Rosner, \prl{65}{90}{2963};
D.C. Kennedy and P. Langacker, \prl{65}{90}{2967}; \prd{44}{91}{1591}.}

\ref\tlimit{P. Langacker, U. Penn preprint UPR-0492T (1991).}

\ref\derujula{A. de R\'ujula, M.B. Gavela, P. Hernandez and E. Mass\'o,
CERN preprint CERN-Th.6272/91, 1991.}

\ref\equivalence{M.S. Chanowitz, M. Golden and H. Georgi,
\prd{36}{87}{1490}; C.P. Burgess and David London, preprint McGill-92/04,
UdeM-LPN-TH-83, 1992.}

\ref\cutoff{C.P. Burgess and David London, preprint McGill-92/05,
UdeM-LPN-TH-84, 1992.}

\ref\weinberg{S. Weinberg, \pr{140}{65}{B516}; in {\it Asymptotic Realms Of
Physics} (Cambridge, 1981).}


\def\smgroup{SU(2)_L\times U(1)_Y}


\rightline{September 1992}
\rightline{UdeM-LPN-TH-104}
\rightline{McGill-92/38}

\title
\centerline{Loop Calculations with}
\centerline{Anomalous Gauge Boson Couplings\footnote{$^\dagger$}
{\baselineskip=0.5\baselineskip\rm Invited talk presented by David London
at the {\it Beyond the Standard Model III} conference, Ottawa, Canada, June
1992}\baselineskip=2.0\baselineskip}
\endtitle

\authors
\centerline{David London${}^a$ and C.P. Burgess${}^b$}
\vskip .15in
\centerline{\it ${}^a$ Laboratoire de Physique Nucl\'eaire, Universit\'e de
Montr\'eal}
\centerline{\it C.P. 6128, Montr\'eal, Qu\'ebec, CANADA, H3C 3J7.}
\vskip .1in
\centerline{\it ${}^b$ Physics Department, McGill University}
\centerline{\it 3600 University St., Montr\'eal, Qu\'ebec, CANADA, H3A 2T8.}
\endauthors

\abstract
Analyses which use loop calculations to put constraints on anomalous
trilinear gauge boson couplings (TGC's) often give bounds which are much
too stringent. The reason has nothing to do with gauge invariance, in
contrast to the recent claims of de R\'ujula et.\ al., since the
lagrangians used in these calculations {\it are} gauge invariant, with the
$\smgroup$ symmetry nonlinearly realized. We trace the true cause of the
problem to the improper interpretation of cutoffs in the calculation. The
point is that the cutoff dependence of a loop integral does not necessarily
reflect the true dependence on the heavy physics scale $M$. We illustrate
that, if done carefully, one finds that the true constraints on anomalous
TGC's are much weaker.
\endabstract
\vfill\eject


One of the most important tasks of LEP200 will be to directly measure the
trilinear gauge boson couplings (TGC's), thereby testing the gauge nature
of the $W$- and $Z$-bosons. Of course it is hoped that these measurements
will reveal the presence of new physics, either through deviations of TGC's
from their standard model predictions, or through the appearance of
anomalous gauge boson vertices. Much work has been done in determining the
precision with which TGC's can be measured, both at LEP200 and at other
facilities.

If one wishes to look at the physics of TGC's, it is convenient to use an
effective lagrangian. In this approach the effects of any unknown new heavy
physics are parametrized through the nonrenormalizable interactions, such
as anomalous TGC's, induced among the light particles. In order to
construct the low-energy effective lagrangian, one needs to specify only
two things -- the low-energy particle content and the symmetries of the
lagrangian. Typically one has three possibilities:
\item{1.} Linearly Realized $\smgroup$: In this formulation the low-energy
particle content includes the standard model Higgs doublet.
\item{2.} Nonlinearly Realized $\smgroup$: This framework is also known
under the general rubric of ``chiral lagrangians''. Here the unknown
symmetry breaking sector at low-energy consists only of the three
Nambu-Goldstone bosons which are eaten to produce the massive $W^\pm$ and
$Z^0$ particles.
\item{3.} Only $U(1)_{em}$ Gauge Invariance: The only bosonic fields
included here are the massive $W^\pm$ and $Z^0$ and the massless photon.
The lagrangian is constrained to respect only Lorentz invariance and
electromagnetic gauge invariance.

Most analyses in the past have used this third formulation in performing
their calculations (although we will return to this point later). The TGC
sector then has the following form for on-shell gauge bosons \LWWV:
\eqa
\label\wwvlag
{\cal L}_{WWV} \sim &
{}~i g_1^V\left(W_{\mu\nu}^\dagger W^\mu V^\nu
           - W^{\mu\nu} W^\dagger_\mu V_\nu \right)
+ i \kappa_V W_\mu^\dagger W_\nu V^{\mu\nu} \eolnn
+& ~i {\lambda_V \over M^2}
             W_{\lambda\mu}^\dagger W^\mu_{~~\nu} V^{\nu\lambda}
- g_4^Z W_\mu^\dagger W_\nu
       \left( \partial^\mu Z^\nu + \partial^\nu Z^\mu\right) + \cdots \eeol
\eeq
where we have included only a subset of all possible TGC's. Here, $V^\mu$
represents either the photon or the $Z$, $W^\mu$ is the $W^-$ field, and
$W_{\mu\nu}=\partial_\mu W_\nu - \partial_\nu W_\mu$ (and similarly for
$V_{\mu\nu}$).

Since we will have to wait several years before these TGC's are measured
directly, it is reasonable to ask if it is possible to put any constraints
on anomalous TGC's now, using current data. Clearly the effects of these
new vertices will only appear at the loop level, and many authors have
calculated the contributions of anomalous TGC's to such loop-induced
processes as the $W$- and $Z$-masses, $\Delta\rho$, $(g-2)_\mu$, and others
\gbloop.

A typical such calculation might go as follows. Consider the contribution
of the above $g_4^Z$ term (which is called the anapole coupling) to the
$\rho$-parameter. A convenient way of representing the effects of new
physics on $\Delta\rho=\rho-1$ is given in terms of the transverse piece of
the gauge boson vacuum polarizations, $\pi_T^{\mu\nu}(q) = g^{\mu\nu}
\pi\left(q^2\right) + ...$ \stu:
\eq
\label\rhoparameter
{\delta \pi_{WW}(0) \over M_W^2} - {\delta \pi_{ZZ}(0) \over M_Z^2}
= \alpha\left(M_Z\right) T.
\eeq
Present constraints on $\Delta\rho$ imply that $T$ satisfies roughly
$\vert T\vert<0.8$ \tlimit. Since the anapole is CP violating, there is a
nonzero effect only when the anapole coupling appears at both vertices of
the vacuum polarization diagrams, as in Fig.~1. Furthermore, since this
coupling is nonrenormalizable, these diagrams will diverge and must be
regularized in some way. The most common regularization method is to use a
cutoff, $\Lambda$. Calculating the diagrams of Fig.~1 in this way, and
keeping only the highest divergence in each case, one obtains
\eqa
\delta\pi_{WW} \left(q^2\right) = &
{}~- { \left(g_4^Z\right)^2 \over 6 \pi^2 }
\thinspace { \Lambda^6 \over M_W^2 M_Z^2 }\eolnn
\delta\pi_{ZZ} \left(q^2\right) = &
{}~- { \left(g_4^Z\right)^2 \over 144 \pi^2 }
\thinspace { \Lambda^4 \over M_W^4 } \thinspace q^2.\eeol
\eeq
If we assume, as is done in the literature, that $\Lambda$ actually
represents the scale of new physics, say 1 TeV, then we can use
Eq.~\rhoparameter\ and the bound on $T$ to put an extremely stringent
constraint on $g_4^Z$:
\eq
\label\overestimate
g_4^Z < 3.5 \times 10^{-4} \left( { 1~{\rm TeV} \over \Lambda } \right)^3.
\eeq

\vskip3truecm
\centerline{Fig.~1: Contribution of anapole anomalous TGC (blob) to gauge
boson propagators.}
\vskip0.3truecm

Recently de R\'ujula and coworkers \derujula\ have made the claim that this
type of analysis is wrong, that it yields constraints which are
considerable overestimates. The cause of this, they say, is that the
lagrangian in Eq.~\wwvlag\ is not gauge invariant under $\smgroup$. This is
a red herring. In fact, the lagrangian in Eq.~\wwvlag\ is equivalent,
term by term, to a chiral lagrangian in which $\smgroup$ is present, but
nonlinearly realized. In other words, although we listed above three
possibilities for a low-energy effective lagrangian, in fact two of them
(the formulations with nonlinearly realized $\smgroup$ and with only
$U(1)_{em}$ gauge invariance) are equivalent. We will not prove this, but
rather refer the reader to Ref.~6 for the details.

It is nevertheless true that results such as Eq.~\overestimate\
overestimate considerably the effects of anomalous TGC's in loops. The real
reason has nothing to do with gauge invariance -- rather it is the improper
use of cutoffs in the calculations of such loop diagrams \cutoff.

Suppose we knew the full theory at scale $M\gg m$, where $m$ represents the
mass scale of some light particle, such as the $W$ or $Z$. And suppose we
now calculate the contribution to a light particle mass, $\delta\mu^2$, as
a function of these two mass scales. The answer will in general have the
following form:
\eq
\label\fullresult
\delta\mu^2 (m,M) = a M^2 + b\thinspace m^2 + c\thinspace \frac{m^4}{M^2} +
\cdots
\eeq
in which the dots represent terms that are suppressed by more than two
powers of $m/M$. Notice that there are no terms of the form $M^4/m^2$.
These are forbidden because only logarithmic infrared divergences are
possible at zero temperature in four dimensions \weinberg. Note also that
the dimensionless coefficients may depend logarithmically on the large mass
ratio $M/m$.

Now suppose that we split the calculation into a ``high-energy'' piece and
a ``low-energy'' piece by choosing a cutoff $\Lambda$ which satisfies
$M\gg\Lambda\gg m$. The contributions from the two pieces might have the
following form:
\eqa
\label\heresult
\delta \mu^2_{\rm he} (m,\Lambda,M) &= a^\prime M^2 + b^\prime \Lambda^2
+ c^\prime {\Lambda^4\over m^2} + \cdots \eolnn
\delta \mu^2_{\rm le} (m,\Lambda,M) &= b^{\prime\prime} \Lambda^2
+ c^{\prime\prime} {\Lambda^4\over m^2} + \cdots \eeol
\eeq
Since the calculation with the cutoff just represents a reorganization of
the full calculation, we must have
\eq
\delta \mu^2(m,M) = \delta \mu^2_{\rm le}(m,\Lambda,M)
+ \delta \mu^2_{\rm he}(m,\Lambda,M),
\eeq
and since the full calculation (Eq.~\fullresult) is independent of
$\Lambda$, it follows that
\eq
a=a^\prime ~,~~~~~~~ b^\prime=-b^{\prime\prime} ~,~~~~~~~
c^\prime=-c^{\prime\prime}.
\eeq
In other words, all quadratic and higher dependence on $\Lambda$ found in
the low-energy calculation is simply cancelled by counterterms coming from
the high-energy piece of the calculation! In this way one sees that there
is no physical significance to terms containing the cutoff $\Lambda$.
Therefore, due to the faulty interpretation of the significance of the
cutoff, results such as that of Eq.~\overestimate\ clearly overestimate the
bounds which can be placed on anomalous TGC's from loop calculations. We
note in passing that the only term which depends strongly on the heavy mass
scale, $a M^2$ in Eq.~\fullresult, cannot be calculated solely within the
low-energy theory -- the coefficient $b^{\prime\prime}$ of Eq.~\heresult\
is completely unrelated to the coefficient of $M^2$ in the full theory.

The point is that cutoffs do not, in general, accurately track the heavy
mass contributions to low-energy processes. The one exception is in the
case of a logarithmic divergence. Suppose there were a term in the full
calculation of the form
\eq
\delta\mu^2 \sim d\thinspace m^2 \log\left({M^2\over m^2}\right).
\eeq
In this case the high-energy and low-energy contributions would have the
form
\eqa
\delta \mu^2_{\rm he} & \sim d^\prime m^2
\log\left({M^2\over \Lambda^2}\right),\eolnn
\delta \mu^2_{\rm le} & \sim d^{\prime\prime} m^2
\log\left({\Lambda^2\over m^2}\right),\eeol
\eeq
so that the cancellation of the $\Lambda$ dependence requires
$d=d^\prime=d^{\prime\prime}$. This is the only case in which the heavy
mass dependence is accurately tracked by the cutoff.

Let us now return to our original example of the anapole contribution to
$\Delta\rho$. Given that the naive use of a cutoff doesn't yield accurate
constraints, how does one get bounds which are physically significant? The
easiest method is not to use cutoffs at all, but to use dimensional
regularization and the decoupling subtraction renormalization scheme. Here,
the divergent contributions to the $W$- and $Z$-propagators are
\eqa
\delta\pi_{WW} \left(q^2\right)\vert_{q^2=0} = &
{}~- { \left(g_4^Z\right)^2 \over 4 \pi^2 } {3 M_W^2 \over 2}
\left[1 + {M_Z^2\over M_W^2} - {M_Z^4\over M_W^4}\right] {2\over \epsilon},
\eolnn
\delta\pi_{ZZ} \left(q^2\right)\vert_{q^2=0} = & ~0, \eeol
\eeq
where $\epsilon=n-4$ in $n$ spacetime dimensions. These divergences
renormalize the bare $T$ parameter, which is also present in the effective
lagrangian:
\eq
\label\mixing
\alpha T(\mu^2) = \alpha T \left({\mu'}^2 \right) - \frac{3}{8\pi^2} \;
\left( g_4^Z \left({\mu'}^2 \right) \right)^2
\; \left[ 1 + \frac{M_Z^2}{M_W^2} - \frac{M_Z^4}{M_W^4}
\right] \; \ln \left( \frac{{\mu'}^2}{\mu^2} \right).
\eeq
This shows how the two operators $T$ and $g_4^Z$ mix when the lagrangian is
renormalized and evolved down from scale $\mu^\prime$ to scale $\mu$ (in
the absence of mass thresholds). Note that, although there may be a strong
quadratic dependence on the heavy mass scale $M$, this is completely
contained in the incalculable initial condition $T\left(M^2 \right)$.

The important point to realize here is that, even in an effective
lagrangian, bare parameters must be renormalized. This seems to have been
overlooked in most previous analyses. Since the effective lagrangian is
nonrenormalizable, in general renormalization will introduce an infinite
number of counterterms. However, this creates no problems, since the
effective lagrangian already contains an infinite number of terms.

We can now use Eq.~\mixing\ to put a bound on $g_4^Z$. Taking
$\mu^\prime\simeq 1$ TeV, $\mu=M_W$, and assuming no accidental
cancellations between the two terms, one finds
\eq
g_4^Z < 0.24.
\eeq
This is 3 orders of magnitude weaker than the bound found using cutoffs
(Eq.~\overestimate)!

In summary, bounds on anomalous trilinear gauge boson couplings are often
significantly overestimated. This is also true for predictions of large
loop-induced effects due to new effective operators. These conclusions are
unrelated to issues of gauge invariance. In fact, any lagrangian which
obeys Lorentz invariance and electromagnetic gauge invariance is
automatically $\smgroup$ gauge invariant, with the $\smgroup\to U(1)_{em}$
symmetry breaking pattern nonlinearly realized. The real source of the
problem is the incorrect use of cutoffs in estimating the effect of the
heavy mass scale in the loops. In general, the cutoff dependence does not
correctly reflect the true dependence on the heavy mass scale, $M$. (The
only exception is in the case of a logarithmic divergence.) An easier way
to do the analysis is to use dimensional regularization, supplemented by
the decoupling subtraction renormalization scheme. This shows explicitly
how operators mix as the effective lagrangian is renormalized. In addition,
one sees that the logarithmic dependence on the heavy mass is simply due to
the renormalization group evolution of the lagrangian down from the scale
$M$ to low energies.

\bigskip
\centerline{\bf Acknowledgments}
\bigskip

This work was supported in part by the Natural Sciences and Engineering
Research Council of Canada, and by FCAR, Qu\'ebec.

\listrefs

\bye